\documentclass[preprint2]{aastex}

\usepackage{rotating}
\usepackage{natbib}
\usepackage{gensymb}
\usepackage{graphicx}
\usepackage{epstopdf}
\usepackage{amsmath}

\newcommand{\teff}{\mbox{$T_{\rm eff}$}}
\newcommand{\logg}{\mbox{$\log g$}}
\newcommand{\vsini}{\mbox{$v \sin i$}}
\newcommand{\mictrb}{\mbox{$\xi_{\rm t}$}}
\newcommand{\mactrb}{\mbox{$v_{\rm mac}$}}

\newcommand{\kms}{\mbox{km\,s$^{-1}$}}

\shorttitle{WASP-135b: a highly irradiated, inflated hot Jupiter orbiting a G5V star}
\shortauthors{Spake et al.}

\begin{document}

\title{WASP-135b: a highly irradiated, inflated hot Jupiter orbiting a G5V star}

\author{J. J. Spake\altaffilmark{1,}\altaffilmark{2},  D. J. A. Brown\altaffilmark{2}, A. P. Doyle\altaffilmark{2}, G. H\'{e}brard\altaffilmark{3,}\altaffilmark{4}, J. McCormac\altaffilmark{2}, D. J. Armstrong\altaffilmark{2,}\altaffilmark{5}, D. Pollacco\altaffilmark{2}, Y. G\'omez Maqueo Chew\altaffilmark{6}, D. R. Anderson\altaffilmark{7}, S. C. C. Barros\altaffilmark{8}, F. Bouchy\altaffilmark{8}, P. Boumis\altaffilmark{9}, G. Bruno\altaffilmark{8}, A. Collier Cameron\altaffilmark{10}, B. Courcol\altaffilmark{8}, G. R. Davies\altaffilmark{11,}\altaffilmark{12},  F. Faedi\altaffilmark{2}, C. Hellier\altaffilmark{7},   J. Kirk\altaffilmark{2}, K. W. F. Lam\altaffilmark{2}, A. Liakos\altaffilmark{9}, T. Louden\altaffilmark{2},  P. F. L. Maxted\altaffilmark{7}, H. P. Osborn\altaffilmark{2}, E. Palle\altaffilmark{13,}\altaffilmark{14}, J. Prieto Arranz\altaffilmark{13,}\altaffilmark{14}, S. Udry\altaffilmark{15}, S. R. Walker\altaffilmark{2}, R. G. West\altaffilmark{2}, P. J. Wheatley\altaffilmark{2}}

\altaffiltext{1}{Astrophysics Group, School of Physics, University of Exeter, Stocker Road, Exeter, EX4 4QL, UK}
\altaffiltext{2}{Department of Physics, University of Warwick, Coventry CV4 7AL, UK}
\altaffiltext{3}{Institut d'Astrophysique de Paris, UMR7095 CNRS, Universit\'e Pierre \& Marie Curie,98bis boulevard Arago, 75014 Paris, France}
\altaffiltext{4}{Observatoire de Haute-Provence, CNRS, Universit\'e d'Aix-Marseille, 04870 Saint-Michel-l'Observatoire, France}
\altaffiltext{5}{ARC, School of Mathematics \& Physics, Queen?s University Belfast, University Road, Belfast BT7 1NN, UK}
\altaffiltext{6}{Instituto de Astronom\'ia, UNAM, 04510, M\'exico, D.F., M\'exico }
\altaffiltext{7}{Astrophysics Group, Keele University, Staffordshire ST5 5BG, UK}
\altaffiltext{8}{Aix Marseille Universit\'e, CNRS, LAM (Laboratoire d'Astrophysique de Marseille) UMR 7326, 13388, Marseille, France}
\altaffiltext{9}{IAASARS, National Observatory of Athens, GR-15236 Penteli, Greece}
\altaffiltext{10}{SUPA, School of Physics \& Astronomy, University of St. Andrews, North Haugh, St. Andrews, Fife KY16 9SS, UK}
\altaffiltext{11}{School of Physics and Astronomy, University of Birmingham, Birmingham, B15 2TT, UK}
\altaffiltext{12}{Stellar Astrophysics Centre (SAC), Department of Physics and Astronomy, Aarhus University, Ny Munkegade 120, DK-8000 Aarhus C, Denmark}
\altaffiltext{13}{Instituto de Astrofisica de Canarias, Via Lactea s/n 38200, La Laguna, Spain}
\altaffiltext{14}{Departamento de Astrofísica, Universidad de La Laguna (ULL), La Laguna, 38205, Tenerife, Spain}
\altaffiltext{15}{Observatoire de Gen\'{e}ve, Universit\'{e} de Gen\'{e}ve, Chemin des mailletes 51, 1290 Sauverny, Switzerland}
\email{jspake@astro.ex.ac.uk}

\begin{abstract}
We report the discovery of a new transiting planet from the WASP survey.  WASP-135b is a hot Jupiter with a radius of 1.30$\pm 0.09$ $R_{\textrm{Jup}}$, a mass of 1.90$\pm$0.08 $M_{\textrm{Jup}}$  and an orbital period of 1.401 days.  
Its host is a Sun-like star, with a G5 spectral type and a mass and radius of 0.98$\pm$0.06 $M_{\sun}$ and 0.96$\pm$0.05 $R_{\sun}$ respectively.
The proximity of the planet to its host means that WASP-135b receives high levels of insolation, which may be the cause of its inflated radius.  Additionally, we find weak evidence of a transfer of angular momentum from the planet to its star.

\end{abstract}


\section{Introduction}
The discovery of the first hot Jupiter \citep{MayorQueloz1995} was surprising since no Solar System analogue existed, and gas giants are not expected to form so close to their parent star.  They are an intriguing class of exoplanets, and those that transit are important since they remain the only planets for which accurate measurements of both mass and radius can be made.  Additionally, they are excellent targets for studying the atmospheric compositions of exoplanets, through transit spectroscopy. 

There is still much that is not entirely understood about hot Jupiters, such as their formation and dynamical evolution mechanisms, and the fact that many appear to have inflated radii (see \cite{FortneyNettleman2010}, \cite{Baraffe2010} and \cite{SpiegelBurrows2013}).  It has been suggested that the unexpectedly large radii of some hot Jupiters can be explained by the injection of some of the received stellar energy into the planetary interior  \citep{Showman2002}.  Indeed there is growing evidence of a positive correlation between levels of incident flux and giant planet radius, for example in \cite{Enoch2011}, \cite{DemorySeager2011}, and  \cite{Weiss2013}. However, the exact mechanisms which transport the energy into the interior of planets remain disputed.  Proposed theories include Ohmic heating \citep{Batygin2010} and kinetic heating \citep{GuillotShowman2002}, among others.

Some, such as \cite{Lanza2010}, suggest that for hot Jupiter host stars, a gyrochronological age significantly younger than age estimates found by other means is evidence of a transfer of angular momentum from the massive, close-in planet to the star, causing tidal `spin-up'.  This process would hinder the normal stellar spin-down mechanisms, leaving the star rotating faster, and appearing younger than expected from gyrochronological models.  Some hints of this effect have been seen, for example \cite{Pont2009} finds empirical evidence of faster rotation in stars with closer and more massive planets, and \cite{Brown2014}  and \cite{Maxted2015} both find a tendency for planet hosting stars to have younger gyrochronological ages than their isochronal age estimates. 

A larger sample of exoplanets is needed in order to investigate the possible spin-up effect of hot Jupiters on their host stars, and to better understand the relationship between incident stellar flux, planet mass and planet radius.  These are some of the reasons that the discovery of transiting hot Jupiters continues to be important.  We present here the discovery of one such planet, WASP-135b.

\section{Observations}
\subsection{SuperWASP-N Photometry}
WASP-135 (1SWASPJ174908.40+295244.9) is a G5V star with $V=13.28$ and $B-V=0.97$ \citep{Zacharias2013}.  It was observed by SuperWASP-N at the Observatorio del Roque de los Muchachos on La Palma, Spain from 2004 May 05 to 2010 August 24, and 34699 observations were made in total. The instrument consists of 8 cameras, with Canon 200-mm f/1.8 lenses and $2048\times 2048$ e2V CCD detectors, each with a field-of-view of $7.8\degree\times 7.8\degree$ and a plate scale of 13".7 pixel$^{-1}$, and broadband V+R filters. The WASP survey and data reduction procedures are discussed in detail in \cite{Pollacco2006}.  

An exoplanet candidate with a period of 1.401 days was identified using transit detection methods described in \cite{CollierCameron2006} and \cite{Pollacco2008}.  The top panel of Figure \ref{photometry} shows the WASP photometry, folded on the best-fitting orbital period.  Additional transit signals were searched for in the WASP data using the method described in \cite{Smith2009} for planet candidates with periods up to 40 days but none were found.  Although the WASP data spans over 6.5 years, systematic noise limits the search for periodicity on very long timescales. The WASP data was also searched for photometric variability as could be caused by magnetic activity and stellar rotation, as described in \cite{Maxted2011}.  No significant signals were found above the 3-mmag level.

Before analysis, the timings for all data including WASP photometry were converted to BJD$_{\text{TDB}}$ using the method of \cite{Eastman2010}.
\begin{figure}[h!]
	\centering
    \includegraphics[width=0.5\textwidth]{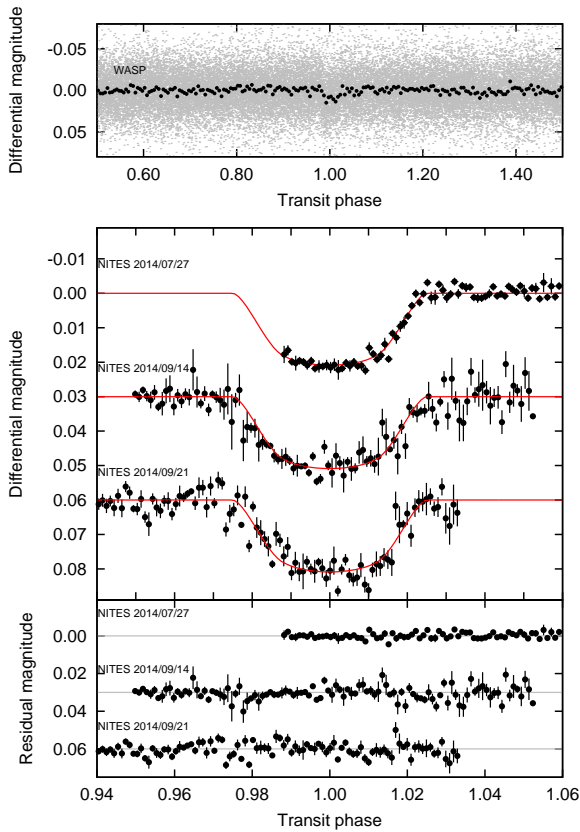} 
    \caption{Top panel: WASP data (through broadband V+R filters) phase-folded on period 1.40138 days, with over-plotted bins of width $\sim$7 minutes.  Lower panels: binned follow-up photometry from NITES (no filter used) with residuals, bin widths $\sim$3 minutes, over plotted with the best fitting transiting planet model using the formalism of \cite{MandelAgol2002}.}
     \label{photometry}
\end{figure}

\subsection{SOPHIE Radial Velocities}
Spectroscopic follow-up of WASP 135 was conducted with the SOPHIE spectrograph mounted on the 1.93m telescope of the Observatoire de Haute-Provence, France.  More details about the instrument can be found in \cite{Bouchy2009}. 

Measurements were made of the radial velocity of the star at many points throughout the proposed orbital phase. In the case of a planetary system, the signal should describe a Keplerian motion of the planet host around the system's centre of mass, yielding the planetary mass and orbital eccentricity, among other parameters.  The radial velocity measurements also allow us to discard astrophysical false positives such as blended stellar binaries.

SOPHIE was used in High-Efficiency mode (HE) with a resolving power $R=40000$ to increase the throughput for this faint target. The typical exposure times were 30 minutes, but they were slightly adjusted as a function of the weather conditions to keep the signal-to-noise ratio at $\sim$30. The spectra were extracted using the SOPHIE pipeline, and the radial velocities were measured from the weighted cross-correlation with a numerical mask \citep{Baranne1996, Pepe2002}. We adjusted the number of spectral orders used in the cross-correlation to reduce the dispersion of the measurements. Indeed, some spectral domains are noisy (especially in the blue part of the spectra), so using them degrades the accuracy of the radial velocity measurement.

The error bars on the radial velocities were computed from the cross-correlation function (CCF) using the method presented by \cite{Boisse2010}. 
We estimated the moonlight contamination using the second SOPHIE fiber aperture, which is targeted on the sky while the first aperture points toward the target. We found RV shifts due to Moonlight contamination below 10 ms$^{-1}$ in all cases, and concluded that there is no significant Moon pollution in our data.  Radial velocities measured using different stellar masks (F0, G2, K0, or K5) produce variations with similar amplitudes, so there is no evidence that their variations could be explained by blend scenarios caused by stars of different spectral types. 

\begin{figure}[ht]
	\centering
	\begin{turn}{270}
    \includegraphics[width=0.35\textwidth]{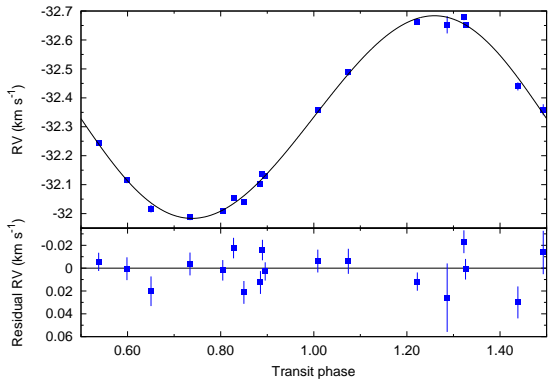}
    \end{turn}
      \caption{All SOPHIE radial velocity data, with the best fitting Keplerian model over plotted. The variations in phase with the transit photometry suggest a sub-stellar companion of $\sim 2M_{\textrm{Jup}}$.}
    \label{RVfig}
\end{figure}

In total 21 observations were made between 2014 May 30 and 2014 October 02.  These are reported in Table \ref{wasp135-RVs} and displayed in Figure \ref{RVfig}, together with the Keplerian fit and the residuals. They show variations in phase with the SuperWASP transit ephemeris and with a semi-amplitude of a few hundred ms$^{-1}$, implying a companion of $\sim 2M_{\textrm{Jup}}$.  

\begin{figure}[h!]
	\centering
	\begin{turn}{270}	
	\includegraphics[width = 0.35\textwidth]{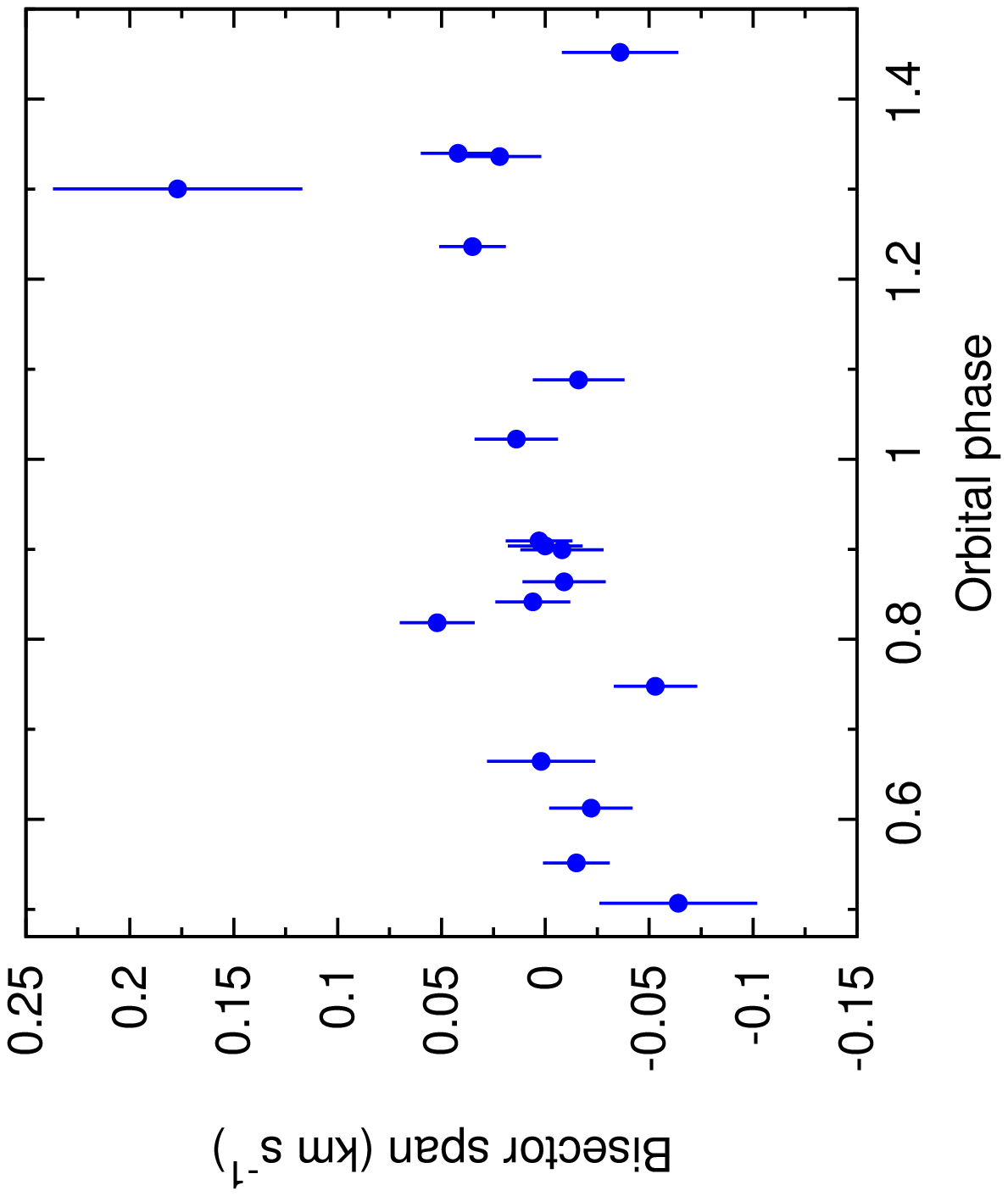}
	\end{turn}\\
	\begin{turn}{270}	
	\includegraphics[width = 0.35\textwidth]{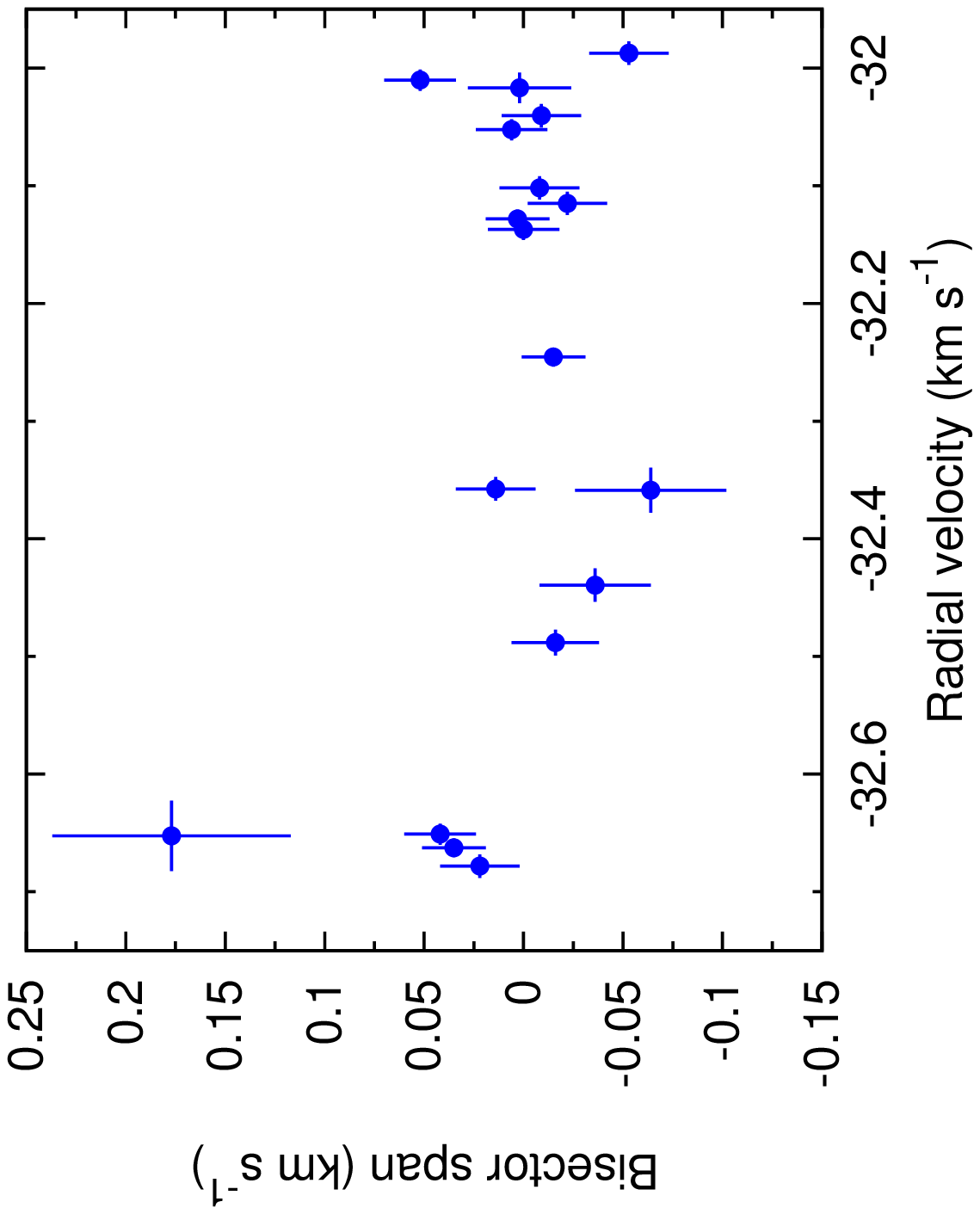}
	\end{turn}
	\caption{CCF bisector spans plotted against orbital phase and radial velocity.}
	\label{bisectors}
\end{figure}

The bisector spans as defined by \cite{TonerGray1988} were measured, and applied to the cross-correlation function as in \cite{Queloz2001}.  They show neither variations nor trends as a function of radial velocity or orbital phase (Fig. \ref{bisectors}).  Additionally, the RMS of the bisector spans is 0.05 kms$^{-1}$, which is of the order of the radial velocity errors.  This reinforces the conclusion that the radial-velocity variations are not caused by spectral-line profile changes attributable to blends or stellar activity. 

Two radial velocity measurements were not included in the final analysis and hence do not appear in the figure, although they are included Table \ref{wasp135-RVs} for completeness.  One (BJD = 2456923.4) was particularly inaccurate due to poor weather conditions.  The other (BJD = 2456813.4) is an outlier, and is off by about 100 m/s from the Keplerian model. We looked for possible causes of this shift, including cosmic rays on the detector, abnormally high drift seen in the wavelength calibration, concern with the CCD reading mode, standard stars observed at the same epoch, sporadic stellar activity events linked to CaII chromospheric emission; we found nothing abnormal. However, a more careful inspection of the CCF shows a small deformation of its profile; it is also seen to have a bisector value slightly lower than values at other epochs. Even if the cause of the CCF deformation is not well understood, it explains the abnormal radial velocity at this epoch. This reinforces its interpretation as an instrumental outlier instead of a signature of an additional companion in the system, in agreement with the fact that other RV measurements secured at similar epochs or similar orbital phase do not show such shifts from the 1-planet Keplerian model.

We thus conclude that our target harbours a transiting giant planet, which we designate WASP-135b hereafter.

\begin{table}[h]
\centering
\caption{Radial velocity measurements of WASP-135 from SOPHIE, with bisector spans from spectral line profiles.}
\begin{tabular}{cccc} \hline
BJD  & RV  & $\sigma _{RV}$  & BS \\  
$-240000$ & (kms$^{-1}$) & (kms$^{-1}$) & (kms$^{-1}$)\\ \hline
56808.4537\phantom{$^{a}$}         &   $-32.017$      &  0.013   &\phantom{$-$}$ 0.002$\\  
56809.3954\phantom{$^{a}$}         &   $-32.678$      &  0.010   &\phantom{$-$}$ 0.022$\\  
56810.4489\phantom{$^{a}$}         &   $-32.488$      &  0.011   &$-0.016$\\  
56811.4723\phantom{$^{a}$}         &   $-32.010$      &  0.009   &\phantom{$-$}$ 0.052$\\  
56812.4370\phantom{$^{a}$}         &   $-32.359$      &  0.019   &$-0.064$\\  
56813.4209$^{a}$                   &   $-32.535$      &  0.015   &$-0.163$\\  
56814.3886\phantom{$^{a}$}         &   $-32.102$      &  0.010   &$-0.008$\\  
56828.4083\phantom{$^{a}$}         &   $-32.137$      &  0.009   &$-0.000$\\  
56829.4019\phantom{$^{a}$}         &   $-32.115$      &  0.010   &$-0.022$\\  
56830.4209\phantom{$^{a}$}         &   $-32.651$      &  0.009   &\phantom{$-$}$ 0.042$\\  
56856.3802\phantom{$^{a}$}         &   $-32.040$      &  0.010   &$-0.009$\\  
56858.3932\phantom{$^{a}$}         &   $-32.653$      &  0.030   &\phantom{$-$}$ 0.177$\\  
56868.4157\phantom{$^{a}$}         &   $-32.439$      &  0.014   &$-0.036$\\  
56870.3629\phantom{$^{a}$}         &   $-32.052$      &  0.009   &\phantom{$-$}$ 0.006$\\  
56897.4126\phantom{$^{a}$}         &   $-32.532$      &  0.022   &\phantom{$-$}$ 0.004$\\  
56899.3859\phantom{$^{a}$}         &   $-32.246$      &  0.008   &$-0.015$\\  
56900.3450\phantom{$^{a}$}         &   $-32.663$      &  0.008   &\phantom{$-$}$ 0.035$\\  
56922.3090\phantom{$^{a}$}         &   $-32.128$      &  0.008   &\phantom{$-$}$ 0.003$\\  
56923.4023$^{b}$                   &   $-32.143$      &  0.063   &$-0.338$\\  
56932.2772\phantom{$^{a}$}         &   $-32.358$      &  0.010   &\phantom{$-$}$ 0.014$\\  
56933.2939\phantom{$^{a}$}         &   $-31.987$      &  0.010   &$-0.053$\\  
\hline 
\\
\end{tabular}
\label{wasp135-RVs}
\newline \begin{footnotesize}
{\bf Notes:} $a$: this measurement was a significant outlier, most likely due to instrumental effects
$b$: poor weather conditions meant that this measurement was particularly inaccurate.  Neither were used in the MCMC analysis.
\end{footnotesize}
\end{table}

\subsection{NITES Photometry}
\label{photometry}
Three nights of follow-up observations were carried out with the Near Infra-red Transiting ExoplanetS (NITES) telescope on 2014 July 27, 2014 September 21 and 2014 September 28.  NITES is a 0.4m, f/10 Meade telescope located at the Observatorio del Roque de los Muchachos on La Palma, equipped with a camera containing a $1024\times1024$ pixel CCD manufactured by e2v, and no filter.  The field of view of the instrument is $11.3\arcmin \times 11.3\arcmin$ and it has a plate scale of 0.66\arcsec pixel$^{-1}$. Autoguiding corrections are calculated directly from the science images using the DONUTS algorithm described in \cite{McCormac2013}, which allows autoguiding on defocussed target stars.  NITES is described in more detail in \cite{McCormac2014}.  

A clear transit was observed in each of the 3 follow-up nights, however data acquisition began after ingress on 2014 July 07.  The middle panel of Figure \ref{photometry} shows the three lightcurves from NITES, in bins approximately 3 minutes wide.  The residuals are shown in the bottom panel.  Although the beginning of some of the observations were taken at higher airmass (and consequently show larger scatter), with the simultaneous analysis of all three light curves we are able to constrain time and shape of the ingress.

The smaller plate scale of the NITES instrument showed the presence of a brighter star on the edge of the large aperture used in the WASP photometry, and 6 fainter sources inside the aperture.  This dilution meant that the transit depth in the WASP data was smaller than that found from NITES, by a factor of 2.7.  Because of this, the WASP data was used to constrain the ephemeris timing only and not the transit depth. This is discussed further in Section \ref{MCMCsec}.

Standard data reduction was performed using PyRAF and the routines in IRAF, and DAOPHOT \citep{Stetson1987} was used for aperture photometry using 6 comparison stars.  Since the stars were defocussed initial apertures were placed manually, then shifted for each science image using the DONUTS routine \citep{McCormac2013}.

\section{Stellar Parameters}
\subsection{Spectral Analysis}
The individual SOPHIE spectra of WASP-135 were co-added to produce a
single spectrum with a S/N of around 50 at a wavelength of 5500 \AA. The standard pipeline
reduction products were used in the analysis. The analysis was performed using
the methods given in \cite{2013MNRAS.428.3164D}. The effective temperature (\teff) was determined from the excitation balance of the Fe~{\sc i} lines. The Na~{\sc i} D lines and the ionisation balance of Fe~{\sc i} and Fe~{\sc ii} were used as surface gravity (\logg) diagnostics. The parameters obtained from the analysis are listed in
Table~\ref{wasp135-params}. The iron abundance was determined from
equivalent width measurements of several unblended lines. A value for
microturbulence (\mictrb) was determined from Fe~{\sc i} using the method of
\cite{1984A&A...134..189M}. The quoted error estimates include that given by the uncertainties in \teff\ and \logg, as well as the scatter due to measurement and atomic data uncertainties. Interstellar Na D lines are present in the spectra with an equivalent widths of $\sim$0.06\AA, indicating an extinction of $E(B-V)$ = 0.017 $\pm$ 0.003 using the calibration of
\cite{1997A&A...318..269M}.  Emission is detected in the Ca~{\sc ii} H and K lines; however, for a star of this magnitude the SOPHIE spectra are not good enough to estimate a reliable activity index.

The projected stellar rotation velocity (\vsini) of $4.67 \pm 0.89 \kms$ was determined by fitting the
profiles of several unblended Fe~{\sc i} lines. This agrees with the \vsini  ~value of $4 \pm 1~ \kms$ obtained from the CCF using the calibration by \cite{Boisse2010}.  A value for macroturbulence (\mactrb) of 2.80 $\pm$ 0.73 {\kms} was determined from the calibration of \cite{2014MNRAS.444.3592D}. An instrumental FWHM of 0.15 $\pm$ 0.01~{\AA} was determined from the telluric lines around 6300\AA.

\begin{table}[h]
\centering
\caption{Stellar parameters of WASP-135 from spectroscopic analysis.}
\begin{tabular}{lr} \hline
Parameter  & Value \\ \hline
RA		   &   17$^{h}$49$^{m}$08$^{s}$.4\\
DEC		   &   +29\degree 52'45.0"\\
\teff      &   5680 $\pm$ 60 K \\
\logg      &   4.50 $\pm$ 0.10 \\
\vsini     &   4.67 $\pm$ 0.89 \kms \\
\mactrb    &   2.80 $\pm$ 0.73 \kms \\
\mictrb    &   0.70 $\pm$ 0.10 \kms \\
{[Fe/H]}   &   0.02 $\pm$ 0.13	 \\
$\log A$(Li)  & 2.01 $\pm$ 0.06 \\
Mass       &   1.01 $\pm$ 0.07 M$_{\sun}$ \\
Radius     &   0.93 $\pm$ 0.12 R$_{\sun}$ \\
Sp. Type   &   G5 \\
Distance   &   300 $\pm$ 45 pc \\ \hline 
\\
\end{tabular}
\label{wasp135-params}
\newline \begin{footnotesize}
 {\bf Note:} Mass and Radius estimate using the
\cite{2010A&ARv..18...67T} calibration. Spectral Type estimated from \teff\
using the table in \cite{Gray2008}. Abundances are relative to the solar values obtained by \cite{2009ARA&A..47..481A}
\end{footnotesize}
\end{table}

\subsection{Stellar Age Estimates}
Lithium is detected in the spectra of WASP-135, with an equivalent width of 39~m\AA, corresponding to an abundance of $\log A$(Li) = 2.01 $\pm$ 0.06. This implies a stellar age of 0.60$^{+1.40}_{-0.35}$ Gyr \citep{2005A&A...442..615S}.

The stellar rotation rate ($P = 10.08 \pm 2.32$~d) implied by the {\vsini} from the spectral analysis gives a gyrochronological age of 0.82$^{+0.41}_{-0.23}$~Gyr using the \cite{Barnes2010} relation.  This is an upper age limit as the {\vsini} gives a minimum rotation rate.

Age estimates were also computed using several different stellar models for isochrone fitting using \teff ~and the stellar density ($\rho_{\ast}$) from the spectral analysis and the MCMC fit respectively.  The method is described in \cite{Brown2014}.  The stellar data were fit to isochrones at the central metallicity of {[Fe/H]}=0.02, and at the upper and lower limits.  These results were combined to get the final estimates of the stellar age and mass.  Three sets of stellar models were used: Padova isochrones as described in \cite{Marigo2008} and \cite{Girardi2010}; Yonsei-Yale isochrones as described in \cite{Demarque2004}; and Dartmouth (DSEP) isochrones as described in \cite{Dotter2008}.  The ages calculated from these models ranged between about 3-6 Gyr, and can be found in Table \ref{wasp135-ages} along with the associated stellar masses found with the models.  The weighted mean and standard error of the three isochronal ages was found to be $4.4\pm 2.5$Gyr.  There is a discrepancy between the two young, fairly consistent ages from gyrochronolgy and Lithium abundances, and the older age estimates from isochrone fitting.

\begin{table}[h]
\centering

\caption{Stellar age estimates for WASP-135.}
\begin{tabular}{lll} \hline
Method  & Age (Gyr)  & Stellar Mass \\ \hline
\\[-8pt]
Padova isochrones		&	4.21$^{+6.08}_{-6.70}$		&	0.99$^{+0.12}_{-0.09}$	\\ [4pt]
Yonsei-Yale isochrones	&	3.10$^{+4.72}_{-2.56}$		&	1.01$^{+0.08}_{-0.08}$	\\	[4pt]
DSEP isochrones 		&	5.96$^{+3.91}_{-4.01}$		&	0.95$^{+0.08}_{-0.07}$	\\[4pt]
Average isochronal age     &      $4.4\pm 2.5$				& 						\\ [4pt]
\parbox[t]{3cm}{\textbf{Li abundance}}&	\textbf{0.60$^{+1.40}_{-0.35}$}		& \\ [4pt]
\parbox[t]{3cm}{Gyrochronology\\(upper limit)} &	0.82$^{+0.41}_{-0.23}$	&	\\ [4pt] \hline
\end{tabular}
\label{wasp135-ages}
\newline 
\begin{footnotesize}
{\bf Note:} Padova isochrones described in \cite{Marigo2008} and \cite{Girardi2010}; Yonsei-Yale isochrones described in \cite{Demarque2004}; Dartmouth (DSEP) isochrones described in \cite{Dotter2008}; Gyrochronological age from \cite{Barnes2010} relation; Li abundance age estimated from \cite{2005A&A...442..615S}.
\end{footnotesize}
\end{table}

\section{MCMC System Parameters}
\label{MCMCsec}
All photometry data from WASP, along with the 3 nights of NITES data and 19 radial velocity measurements from SOPHIE were used in a Markov chain Monte Carlo (MCMC) fit.  However, because the photometry from WASP was diluted by other stars inside and on the edge of the aperture, the transit depth in this data set was smaller than that found with NITES by a factor of 2.7.  The in-transit WASP data was therefore corrected by this factor before the MCMC fitting, as has been done for other WASP planets, for example in \cite{2012MNRAS.422.1988A}.  This allowed the WASP data to constrain the ephemeris timing without affecting the transit depth.

The MCMC simultaneously fits a Keplerian orbital solution to the radial velocity data, and the \cite{MandelAgol2002} model to the photometric transits.  
Limb-darkening was accounted for using a non-linear model with four coefficients, which were found by interpolating from the R-band table of \cite{Claret2000} and \cite{Claret2004}, which is appropriate for the NITES photometry.

We used an updated version of the code described in \cite{CollierCameron2007}, \cite{Pollacco2008} and \cite{Brown2012}, to fit for the following parameters: the epoch of mid-transit, T$_{0}$; the orbital period, P; the total transit duration, T$_{14}$; the ingress/egress duration, T$_{12}$=T$_{34}$; the planet to star area ratio, $R^{2}_{P}/R^{2}_{\ast}$; the impact parameter for a circular orbit, b; \teff; [Fe/H], and the RV semi-amplitude, K. We applied priors on \vsini, \teff ~and [Fe/H] using the values listed in Table \ref{wasp135-params}, and a 0.025 kms$^{-1}$ stellar jitter term was included in the fit.
The \cite{LucySweeney} F-test was used to test for eccentricity in the Keplerian orbit.  When the orbit was not forced to be circular in the MCMC fit, an eccentricity of 0.038$\pm$0.011 was found.  According to the F-test the probability that this eccentricity could be found by chance is 0.059, therefore a circular model is adopted in the final MCMC analysis, as recommended in \cite{2012MNRAS.422.1988A}.  The resulting system parameters from this analysis are presented in Table \ref{MCMC_params}.

\begin{table}[h]
\centering
\caption{MCMC system parameters for WASP-135}
\begin{tabular}{lr} \hline
Parameter  & Value \\ \hline
P (d)			&	1.4013794 $\pm$ 0.0000008			\\
$T_{c}~ \textrm{BJD}_{\textrm{TDB}}$	&	2455230.9902 $\pm$ 0.0009			\\
$T_{14}$ (d)		&	0.069 $\pm$ 0.001				\\
$T_{12}=T_{34}$ (d)	&	0.018 $\pm$ 0.002				\\
$\Delta F=R^{2}_{p}/R^{2}_{\ast}$	&	0.0194 $\pm$ 0.0008\\
$\textit{b}$			&	0.80 $\pm$ 0.03\\
$\textit{i}$($\degree$)		&	82.0 $\pm$ 0.6\\
$K_{1}($km s$^{-1})$		&	0.346 $\pm$ 0.003\\
$\gamma ($km s$^{-1})$		&	$-32.334 \pm 0.002$\\
$\textit{e}$			&	0 (adopted)\\
$M_{\ast}(M_{\sun})$		&	0.98 $\pm$ 0.06\\
$R_{\ast}(R_{\sun})$		&	0.96 $\pm$ 0.05\\
log~$g_{\ast}$ (cgs)		&	4.47 $\pm$ 0.03\\
$\rho_{\ast} (\rho_{\sun})$	&	1.12 $\pm$ 0.15\\
\teff (K)			&	5675 $\pm$ 60\\
$M_{P}(M_{\textrm{Jup}})$	&	1.90 $\pm$ 0.08\\
$R_{P}(R_{\textrm{Jup}})$	&	1.30 $\pm$ 0.09\\
log~$g_{P}$ (cgs)		&	3.41 $\pm$ 0.05\\
$\rho_{P} (\rho_{J})$		&	0.87 $\pm$ 0.17\\
$a$ (AU)			&	0.0243 $\pm$ 0.0005\\
\hline 
\end{tabular}
\label{MCMC_params}
\newline \begin{footnotesize}
Notes: Errors are $1\sigma$; limb-darkening coefficients are: a$_{1}=0.664$, a$_{2}=-0.378$, a$_{3}=0.945$, a$_{4}=-0.470$; \textbf{stellar mass and radius are from MCMC analysis of the photometry and RV data, after applying priors from stellar spectral analysis.}\\ 
\end{footnotesize}
\end{table}

\section{Discussion} 
With a large radius of 1.30 $\pm$ 0.09 $R_{\textrm{Jup}}$, WASP-135b appears to be a bloated hot Jupiter. It is a highly irradiated exoplanet, receiving an insolation of $1.98\pm0.24 \times 10^{9}$ erg s$^{-1}$ cm$^{-2}$, which is an order of magnitude greater than the cut off point found by \cite{DemorySeager2011} of $\sim$ $2\times 10^{8}$ erg s$^{-1}$ cm$^{-2}$. They find that giant planets receiving more insolation than this limit have inflated radii.  Additionally, \cite{Weiss2013} find strong evidence for the dependence of giant planet radius on incident flux.  Their empirical relation for planet radius, mass, and incident flux predicts a radius of 1.30 $\pm$ 0.07 $R_{\textrm{Jup}}$ for WASP-135b, which fully agrees with the value measured in this work.

The weighted mean and standard error of the three isochronal ages for WASP-135 presented here (see table \ref{wasp135-ages}) is $4.4\pm 2.5$ Gyr, whilst the gyrochronological age upper limit is 0.82$^{+0.41}_{-0.23}$ Gyr.  This age discrepancy could be evidence of tidal spin-up of the star, due to its massive, short period planet.  On the other hand, lithium is detected in the stellar spectrum, and the age estimate of 0.60$^{+1.40}_{-0.35}$ Gyr  from its abundance measurement lies below the upper age limit from gyrochonological models. Since the errors on the isochronal ages are large, and the gyrochronological and lithium ages agree well, the evidence for stellar spin-up appears weak, suggesting that WASP-135 is likely a relatively young star.  Of the 146 exoplanets with masses greater than 0.1 $M_{\textrm{Jup}}$, orbital distances less than 0.1 AU, and published estimates for their stellar host age, only 7 orbit stars younger than 0.8 Gyr\footnote{exoplanetarchive.ipac.caltech.edu}.

\acknowledgments
The authors would like to thank the anonymous referee for their valuable comments which greatly improved this paper.  
The WASP Consortium consists of representatives from the Universities of Keele, Leicester, The Open University, Queens University Belfast, St Andrews and Warwick, along with the Isaac Newton Group (La Palma) and the Instituto de Astrofisica de Canarias (Tenerife).  SuperWASP-North is hosted by the Isaac Newton Group and the Instituto de Astrofisica de Canarias; we gratefully acknowledge their ongoing support and assistance.  The SuperWASP cameras are operated with funds available from Consortium Universities and the STFC.  The SOPHIE observations were made possible thanks to OPTICON funding allocations 2014A/043 (PI Faedi) and 2014B/029 (PI G\'omez Maqueo Chew). 
YGMC has been partially funded by {\it Programa UNAM-DGAPA-PAPIIT IA-103215}.  This research has made use of NASA's Astrophysics Data System Bibliographic Services.

{\it Facilities:} \facility{SuperWASP}, \facility{OHP:1.93m (SOPHIE)}, \facility{NITES}.

\bibliographystyle{aa}
\bibliography{references}



\end{document}